\documentclass{aa}  

\usepackage{graphicx}
\usepackage{amsmath}
\usepackage{txfonts}

\begin{document}

\title{Comparing the emission spectra of U and Th hollow cathode lamps and a new U line-list}

   \author{L. F. Sarmiento
          \inst{1}
          \fnmsep\thanks{\email{sarmiento.luisfernando@gmail.com.}}
          \and
          A. Reiners\inst{1}
          \and
          P. Huke\inst{1}
          \and
          F. F. Bauer\inst{1}
          \and
          E. W. Guenter{\inst{2}$^{,}$\inst{3}}
          \and
          U. Seemann\inst{1}
          \and
          U. Wolter\inst{4}
}
   \institute{Institut f\"ur Astrophysik, Georg-August-Universit\"at G\"ottingen, Friedrich-Hund-Platz 1, 37077 G\"ottingen, Germany\\
   \and
   Th\"uringer Landessternwarte Tautenburg, Sternwarte 5, 07778 Tautenburg, Germany\\
   \and
   Instituto de Astrof\'isica de Canarias (IAC), 38205 La Laguna, Tenerife, Spain\\
   \and
   Hamburger Sternwarte, Universit\"at Hamburg, Gojenbergsweg 112, 21029 Hamburg, Germany\\}
   \date{Received , 22 February 2018; accepted , 5 June 2018}

% \abstract{}{}{}{}{} 
% 5 {} token are mandatory
 
  \abstract
  {Thorium hollow cathode lamps (HCLs) are used as frequency calibrators for many high resolution astronomical spectrographs, some of which aim for Doppler precision at the 1 m/s level.}
  {We aim to determine the most suitable combination of elements (Th or U, Ar or Ne) for wavelength calibration of astronomical spectrographs, to characterize differences between similar HCLs, and to provide a new U line-list.}
  {We record high resolution spectra of different HCLs using a Fourier transform spectrograph: (i) U-Ne, U-Ar, Th-Ne, and Th-Ar lamps in the spectral range from 500 to 1000~nm and U-Ne and U-Ar from 1000 to 1700~nm; (ii) we systematically compare the number of emission lines and the line intensity ratio for a set of 12 U-Ne HCLs; and (iii) we record a master spectrum of U-Ne to create a new U line-list.}  
  {Uranium lamps show more lines suitable for calibration than Th lamps from 500 to 1000~nm. The filling gas of the lamps significantly affects their performance because Ar and Ne lines contaminate different spectral regions. We find differences (up to 88\%) in the line intensity of U lines in different lamps from the same batch. We find 8239 isolated lines between 500 and 1700~nm\thanks{Table~\ref{line_list_vis} is only available in electronic form at the CDS via anonymous ftp to cdsarc.u-strasbg.fr (130.79.128.5) or via http://cdsweb.u-strasbg.fr/cgi-bin/qcat?J/A+A/} that we attribute to U, 3379 of which were not contained in earlier line-lists.}
  {We suggest using a combination of U-Ne and U-Ar lamps to wavelength-calibrate astronomical spectrographs up to 1~$\mu$\textit{m}. From 1~$\mu$\textit{m} to 1.7~$\mu$\textit{m}, U-Ne shows better properties. The differences in line strength between different HCLs underline the importance of characterizing HCLs in the laboratory. The new 3379 U lines can significantly improve the radial velocity precision of astronomical spectrographs.}

   \maketitle
%
%________________________________________________________________

\section{Introduction}
\label{Int}
Thorium hollow cathode lamps (HCLs) are the \textit{de facto} standard for wavelength calibration of astronomical high resolution spectrographs. The high number of spectral features emitted by HCLs from the cathode make these lamps good calibration sources for astronomical spectrographs. Current instrumentation used in searching for exoplanets (e.g. HARPS, \citealp{Mayor2003} or CARMENES, \citealp{Quirrenbach2014}) use HCLs for wavelength calibration and achieve meter per second radial velocity precision. Commercially available HCLs fulfill most of the criteria of an ideal calibrator (\citealp{Stanley55}; \citealp{Kerber2007a}; \citealp{Lovis2007}): they are simple and safe to handle, are relatively inexpensive and readily available auxiliary equipment, and have a relatively long life time and narrow lines while covering a wide spectral range.\\

Since the first HCLs were introduced by \citet{Paschen1916} as a light source for spectroscopic investigations, continuous improvements have been made. The first HCLs contained a vacuum circulating system, described by \citet{Tolansky1947}, to prevent impurities from outgassing, which shortened the life time of the lamps. Further improvements of HCLs have been driven by the requirements of atomic absorption (AA) spectroscopy developed by Walsh in the 1950s. \citet{Dieke52} introduced an activated U getter (consisting of a vaporized reactive metal layer deposited in the inner wall of the glass tube) for Fe cathode-based HCLs. This design prevents outgassing and yields a relatively long life time for sealed HCLs. \citet{Russell1957} presented zirconium or tantalum getters, which allowed HCLs of a wide variety of cathode materials to be sealed. \citet{Jones1960} presented further improvements and detailed studies about the construction and characteristics of the sealed tubes used as a spectroscopic light source.
 With the introduction of the getter and the improvement of the glass tube to seal-off the HCLs, these lamps became commercially available. Since then, HCLs have been widely used in applications where narrow spectral lines are required, for example in the wavelength calibration of high resolution astronomical spectrographs.\\

Current commercial HCLs use this sealed design and contain a metal cathode, a metal anode, and a buffer gas at a defined pressure. The metal cathode can be either made of a single species or of a combination of different elements. The buffer gas must be a nobel gas to avoid a chemical reaction with the cathode material, which would contaminate the spectrum with molecular emission lines. The front window of the glass tube must be transparent in the wavelength region for which it is used. The physical principles of operating HCLs have been studied in great detail by different authors (\citealp{Crosswhite55}; \citealp{Lieberman1994}; \citealp{Huke}). As explained in \citet{Kerber2008}, HCLs are operated by applying a voltage between the cathode and the anode. Therefore, an electric potential is established between them. The negative potential of the cathode accelerates the electrons towards the anode. During this process, their energy increases until they have sufficient energy to ionize the neutral buffer gas 
atoms through inelastic collisions that create a hot plasma. The cations from the plasma accelerate towards the cathode, colliding with its surface at energies high enough to overcome the electronic work function of the metal cathode. Material from the cathode is sputtered at high velocities, usually in the form of unexcited atoms. The sputtering rate depends on the cathode material, the mass of the buffer gas, and the voltage applied between the cathode and the anode. The sputtered material collides with fast electrons ionizing the metal material, and with atoms from the buffer gas, resulting in unstable excited states. The metal and gas atoms relax back to the ground state and therefore emit photons during the process. Therefore HCLs yield spectral features from two different sources: the material of the cathode and the buffer gas. The intensitiy of the lines is dependent on the operational current. Higher operational current yields a higher voltage difference between the cathode and the anode. The cations 
from the 
buffer gas hit the cathode with higher energies, which in turn increases the rate of sputtered material from the cathode \citep{Kerber2008}. Therefore, not only are more metallic lines emitted, but metal lines also grow faster in intensity than buffer-gas-emitted lines.\\

To use HCLs as calibrators for astronomical spectrographs, the wavelength of the spectral lines must be known with high accuracy and precision. Therefore, HCLs are measured in laboratories using Fourier transform spectrographs (FTS), which offer the required precision and accuracy to carry out these interferometric measurements covering a wide spectral range. Great efforts have been made during the last decades to create line-lists for different cathode elements. The Th line-list presented by \citet{Redman2014} (from now on R2014) combines precise FTS measurements of Th-Ar HCLs with seven previous works (\citealp{Giacchetti1974}; \citealp{Zalubas1974}; \citealp{Zalubas1976}; \citealp{Palmer1983}; \citealp{EnglemanJr2003}; \citealp{Lovis2007}; \citealp{Kerber2008}) to calculate accurate Ritz wavelengths.\\

Pure Th cathode lamps have not been manufactured for several years \citep{Fischer2016}. Current Th HCLs contain a cathode made out of Th-oxide, which introduces impurities into the cathode. High resolution measurements of the Th-oxide lamp's spectrum show undesirable spectral features, also known as a "grass" of unidentified emission lines. These features compromise the wavelength calibration of high accuracy radial velocity spectrogaphs. Uranium HCLs have been proposed to be used for high precision wavelength calibration in the near infrared (NIR) spectral range in \citet{Redman2011} (from now on R2011), where a line-list was published as a line-list from 0.85~$\mu$\textit{m} to 4~$\mu$\textit{m}. R2011 complements previous works: \citet{Palmer1980} created a high energy line-list from 3846~\AA\ to 9091~\AA, and \citet{Conway1984} published a line-list from 1.8~$\mu$\textit{m} to 5.5~$\mu$\textit{m}. However, R2011 has a reduced number of lines in the region around 1080~nm (9250~cm$^{-1}$) because 
of the low sensitivity of the 1-m FTS of the Kitt Peak National Solar Observatory in that band. This spectral region overlaps the \textit{Y} band, which is of special interest for astronomical observations of M dwarfs \citep{Reiners2010}.\\

Because pure Th cathode lamps are currently not available, and there are no line-lists available for alternative cathode materials, wavelength calibration of optical spectrographs becomes increasingly difficult. We explore if U can also be an alternative for the established Th cathodes in the wavelength range from $500$ to $1000$~nm (like in the NIR as proposed by \citealp{Redman2011}). Uranium cathode HCLs are made of natural U with isotope abundances ${^{238}\textrm{U}} \sim 99.3$\% and ${^{235}\textrm{U}} \sim 0.7$\% and a few additional isotopes with abundances smaller than 0.01\% \citep{isotopes}. These isotopes may contribute to lines we identify as U lines as well as to the unidentified lines found in our spectra. However, the second most abundant isotope, ${^{235}\textrm{U}}$ is hyperfine-structured. Therefore, the expected line intensities are low and the contribution to the observed U spectra can be neglected \citep{Redman2011}. We devote this work to the performance comparison between the 
currently available Th and U 
lamps and present a uniform U line-list spanning the wavelength range from $500$ to $1700$~nm.\\

In Sect.~2, we explain our experimental setup and our FTS instrument used to take high resolution spectra of the hollow cathode lamps. In Sect.~3, we identify spectral features in the spectra of four HCLs: U-Ne, U-Ar, Th-Ne, and Th-Ar. Then, we match the features with previously published line-lists and determine the emitting element of the lines found in our spectra. In Sect.~4, we discuss the performance of each of the four lamps. In Sect.~5, we briefly discuss the finding of molecular bands in the spectra of the Th-Ar lamp. Because HCLs are replaced on a regular basis in observatories, a similar performance for lamps of the same type is desired. In Sect.~6, we test 12  individual U-Ne lamps to identify how large the differences between lamps of the same production batch can be. Finally, our U line-list is presented in Sect.~7. Section~8 summarizes our results.\\
%__________________________________________________________________

\section{Experimental setup and measurements}
\label{sec:Experimental setup and measurements}

The lamps used in this work are currently used in the Calar Alto high-Resolution search for M dwarfs with Exoearths with Near-infrared and optical Échelle Spectrographs (CARMENES) \citep{Quirrenbach2014} calibration unit. For this purpose 12 identical U-Ne lamps were ordered from \textit{Photron PTY Ltd.}\footnote{The identification of commercial products in this document intends to specify the experimental procedure properly. It is not intended to imply recommendation or endorsement, nor is it intended to imply that the products identified are necessarily the best available for the purpose.}. The HCLs belong to the same batch and have consecutive serial numbers (to ensure that they were built in a similar process). To record high quality spectra and operate the lamps safely, HCLs must be placed in a housing. It was designed and built by the Th\"uringer Landessternwarte Tautenburg workshop.\\

\begin{figure}[h]
 \centering
  \includegraphics[width=\hsize]{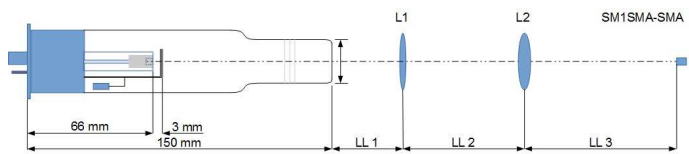}
   \caption{Light from the HCL is coupled into a fibre using a set of optics: two doublets designated by L1 (AC254-125-B-ML from \textit{Thorlabs$^1$}), L2 (AC254-075-B-ML from \textit{Thorlabs$^1$}), and a SM1SMA-SMA fibre adaptor. The text provides further details.}  
      \label{HCL_Opt}
\end{figure}

To feed the light from the glowing cathode into the spectrograph (via a fibre) we used the fore-optics shown in Fig.~\ref{HCL_Opt}. It consists of: L1, a doublet to collimate the light from the cathode located at a distance to the lamp front window (LL1) of 35.5~mm; a second doublet L2, located at a distance to the doublet L1 of 60.3~mm (LL2), which feeds the light into the fibre using a SM1SMA-SMA fibre adaptor. The distance LL3, between the doublet L2 and the fibre adaptor is 75~mm (for further details see the caption of Fig.~\ref{HCL_Opt}). We connected a power-meter to the fibre port to have a simple method for aligning and focusing the light source.\\

We equipped our FTS source chamber with a 8:92 (R:T) \textit{Thorlabs$^1$} pellicle beam splitter and a photodiode. The beam splitter reflected 8\% of the incident light into a photodector, which measured the flux with a frequency of 1~kHz. Simultaneously, we measured the operational voltage of the lamp using a Universal Serial Bus (USB) voltmeter with a frequency of 2~Hz.\\

\begin{figure}[h]
 \centering
  \includegraphics[width=\hsize]{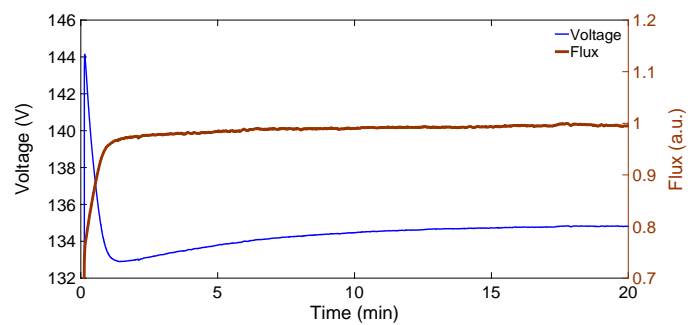}
   \caption{Example of flux (brown) and voltage (blue) during the lamp warm-up time of one of our U-Ne lamps.}
      \label{figure_paper_vevolution}
\end{figure}

Figure~\ref{figure_paper_vevolution} shows the measured voltage (blue) and flux (brown) during $20$~min after switching on one of our lamps. The changes in both parameters, voltage and flux, indicate that the HCL needs a warmup time to produce a stable output. To analyse these changes, we first averaged the voltage and flux measurements every minute. In a second step, we calculated the slope of the relative voltage variation and the relative flux variation of the averaged values. The changes in flux were less than 0.5\% after $7$~min. However, we observed changes in voltage within approximately $14$~min, before the voltage reached a stability of 0.05\%. Therefore we recommend a minimum warm-up time of the \textit{Photron} HCLs of $15$~min to get the most stable output (see Fig.~\ref{voltage_apendix} in Appendix A).\\

We recorded the spectra of the lamps using our IFS125HR FTS. In Table~\ref{table_experiments} (Appendix A), we summarize the resolution, number of scans, and operational current used for the different experiments. To obtain high quality spectra that cover the full CARMENES spectral bandwidth from 500 to 1700~nm (from 20000~cm$^{-1}$ to 5882.4~cm$^{-1}$), we used two different setups. For the following, we call the first setup the visible (VIS) one. We recorded spectra from 500 to 1000~nm (from 20000~cm$^{-1}$ to 10000~cm$^{-1}$) using a quartz beam splitter and a silicon diode detector. The second setup, called the NIR setup, was used to record spectra from 1000 to 1700~nm (from 10000~cm$^{-1}$ to 5882.4~cm$^{-1}$). The NIR setup consists of a calcium fluoride beam splitter and an InGaAs detector.\\

Our FTS uses a frequency-stabilized HeNe laser (at a wavelength of 632.816~nm in air, equivalent to 15802.382~cm$^{-1}$) to obtain the internal wavenumber solution by measuring the optical path difference of the interferometer during the mirrors scanning. A fraction of the laser light is reflected during its path and also enters the detector. Hence, the VIS FTS spectra used for this work are strongly contaminated by the laser in the region between 610 and 655 nm (from 16393.4~cm$^{-1}$ to 15267.2~cm$^{-1}$). Therefore, we are less sensitive to weak spectral lines in the contaminated range, and we can only detect the strongest emission lines in our spectra.\\
%______________________________________________________________ 
\section{Line search, line identification, and spectrum calibration}
\label{sec:Methods for calibration and analysis}

\subsection{Identification of line candidates}
\label{sec:linedet}

The sensitivity of the FTS in combination with the HCL light source yields a low signal to noise ratio (SNR) spectrum in single scans. To increase the signal, we averaged over a high number of individual scans (see Table~\ref{table_experiments} in appendix A).\\

To quantify the noise level of the spectrum, we excluded the points with the highest intensities (10\%) and calculated the standard deviation and the mean of the remaining signal; we defined our noise level as $\mu +\left ( 3\times \sigma \right )$. Next, we defined a wavenumber bandwidth of 0.09~cm$^{-1}$ around each point with intensity higher than the defined noise levels, and selected the data points that have a continuous first derivative centred on the zero crossing and flagged them as "detected line". We excluded the points with intensities lower than the noise level to determine better line parameters. Finally, we fitted a Gaussian curve to each emission line and calculated the centre of the line, the maximum intensity, and the full width at half maximum (FWHM). To estimate the uncertainties of these parameters, we first obtained the statistical uncertainties of the fitted parameters from the covariance matrix. We then applied error propagation to calculate the 
uncertainties of the centre of 
the lines, their maximum intensity, and the FWHM.\\

\subsection{Lines found in the literature}
\label{sec:lineide}

To determine the emitting element for the detected lines, we first looked for a match of these lines with a line in the literature. Next, we confirmed the line identification by analysing the FWHM of the line.\\

We calculated the wavenumber difference between the line position of our detected lines and lines from the R2011 line-list to identify U lines, and the R2014 line-list to identify Th lines. We used the NIST line-list \citep{NIST_ASD} to identify Ne and Ar lines. Detected lines that match a line position listed in the line-lists within three times its estimated position uncertainty were flagged as "confirmed" lines.\\

\begin{figure}[h]
 \centering
  \includegraphics[width=\hsize]{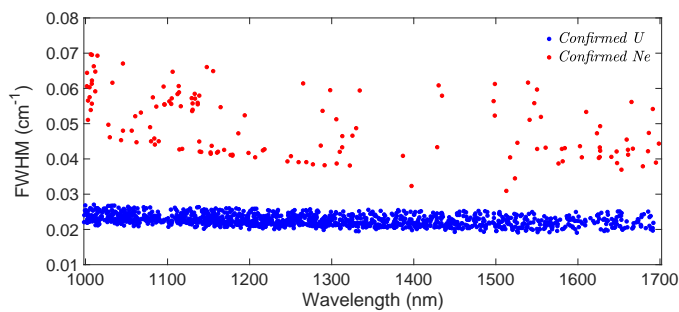}
   \caption{FWHM of confirmed U (blue dots) and confirmed Ne (red dots) lines in the spectrum of one of our U-Ne lamps.}
      \label{FWHM_diff_elements}
\end{figure}

Lines emitted by the metallic material of the cathode (either U or Th) and the lines emitted by the HCL's filling gas (either Ne or Ar) have very different widths (Doppler broadening) because of their different atomic weight (Th:232 u, U:238 u, Ne:20 u, Ar:40 u). Hence, the lines emitted by U or Th are narrower than the lines emitted by Ne or Ar. The possibility that the lines emitted by the buffer gas and the cathode have different FWHM is used to confirm the line identification done by matching the line position. In Fig.~\ref{FWHM_diff_elements}, we show that the FWHMs of metal and gas lines differ significantly between U and Ne lines found in our NIR spectrum.

We fitted the FWHM of the confirmed lines as a function of wavenumber and used this information to determine the emitting element for the remaining detected lines in Sect.~\ref{sec:ident_u_line_cand}. We computed the residuals to the fit and calculated the standard deviation of the residuals. Lines with residuals higher than three times the standard deviation were flagged as blends, while lines with residuals smaller than three times the standard deviation were flagged as "isolated" lines.\\

Before classifying the lines that were not found in the literature, we searched for the traces of additional contaminants that might be present in our recorded spectra. We matched the line position of the unidentified lines with the lines in the line-lists of Ne (in lamps where Ar is used as the filling gas), Ar (in lamps where Ne is used as the filling gas), N (because it is the most abundant gas in the atmosphere), and Zr (because it is used in the lamps tested here to trap molecules sputtered from the cathode) that were downloaded from the National Institute of Standards and Technology (NIST) database \citep{NIST_ASD}. We did not find any significant gas contamination in our measurements. As a result of the first line identification analysis, 87\% of the emission lines remained unidentified in the VIS spectrum, and 50\% in the NIR spectrum (see Table~\ref{table_lineidentification_linelist_vis} and Table~\ref{table_lineidentification_linelist_nir} in Appendix A). In Sect.~\ref{sec:ident_u_line_cand}, we 
attempt to classify the large number of remaining 
unidentified lines to build a list of U line candidates.\\

\subsection{Identification of new U line candidates}
\label{sec:ident_u_line_cand}

Since we found a large number of unidentified lines in our spectra, we are interested in identifying the lines emitted by the metallic cathode. In a first step, we selected those lines from the "unidentified detected lines" that have measured FWHM smaller than the confirmed Ne lines. Next, we checked that those lines are in line with the fitted $\mathrm{FWHM(\nu)}$ relation determined from the confirmed U lines. The lines that matched both criteria were flagged as "U line candidates". To confirm the nature of the U line candidates, we will examine the lines further in Sect.~\ref{sec:U_line_list}. In that section, we will analyse the line intensity, operating the lamps at different currents to confirm the emitter element (as in \citealp{Kerber2007a}).\\

\subsection{Spectrum calibration}
\label{sec:cal}
\subsubsection{Wavenumber calibration}
\label{wave_cali}

Our FTS wavenumber solution is based on an internal frequency-stabilized He-Ne laser. The laser is used to measure the optical path difference of the interferometer during measurements. Any misalignment between the light path of the science light and the internal laser combined with the limited size of the FTS entrance aperture introduces a linear compression or stretch of the instrument wavenumber scale \citep{Griffiths2007}. We corrected for this effect using eight U and seven Th standard lines between 694 and 755~nm \citep{DeGraffenreid2002}. The wavenumber of the standard lines was measured with an accuracy of the order of 10$^{-8}$ (0.0002 cm$^{-1}$), which is one order of magnitude higher than the accuracy of our measured line positions (10$^{-7}$). The wavenumber correction factor was calculated for each standard line \textit{i} as

\begin{center}
        $k_{i}=\dfrac{\tilde{\nu}_{std,i}}{\tilde{\nu}_{obs,i}}-1$,
\end{center}
where $\tilde{\nu}_{std,i}$ is the standard wavenumber of line \textit{i} and $\tilde{\nu}_{obs,i}$ is our measured wavenumber. We calculated the weighted average of the correction factor of all the lines to obtain the spectrum wavenumber correction factor \textit{k}. To convert the FTS wavenumber scale into an absolute scale, we computed
\begin{center}
        $\tilde{\nu}_{c}=(1+\textit{k})\times {\tilde{\nu}_{FTS}}$,
\end{center}
where $\tilde{\nu}_{FTS}$ is the internal FTS wavenumber scale and $\tilde{\nu}_{c}$ is the corrected wavenumber scale.\\

We obtained $k_{VIS}=(1.4\pm0.1)\times 10^{-7}$ for the spectrum measured in the VIS band. No Th or U standard lines are available similar to the ones of \cite{DeGraffenreid2002} for our NIR setup. Therefore, we used 21 confirmed and isolated, high intensity U lines that we found in the wavelength region where our VIS setup and NIR setup overlap to cross calibrate our two setups. The errors estimated for these lines in the VIS spectrum are of the order of 0.001~cm$^{-1}$. Hence, each single line can be measured with a precision comparable to the expected wavenumber correction (see Fig.~\ref{k_nir}). Using the weighted average of the 21 $k_{i}$, we obtained $k_{NIR}=(-2.85\pm0.9)\times 10^{-7}$. We noted that the higher uncertainty of the $k_{NIR}$ was a result of using the 21 absolute calibrated lines from the VIS setup, which are less accurate than the standards lines of \cite{DeGraffenreid2002} used to calibrate the VIS spectrum.\\ 

\subsubsection{Flux calibration}
\label{flux_cali}

We calibrated the line intensity for the spectral response of the FTS and the fore-optics. For this, we recorded a low resolution spectrum of a calibrated tungsten lamp (with a black body temperature of 3000~K). We applied a moving average to obtain the continuum of the spectrum and divided the continuum curve by the theoretical black body curve of the lamp. We identified the calculated curve with the spectral response of the setup. The line intensity of each line was corrected by dividing the calculated line intensity by the spectral response curve.\\ 

\section{Cathode and buffer gas elements selection}
\label{sec:cathode element selection}

In this section, we characterize U-Ne and U-Ar lamps and compare them with Th-Ne and Th-Ar lamps (from 500 to 1000~nm). Since the number of Th lines is known to drop significantly beyond 1000~nm \citep{Redman2014}, we restricted our analysis for the NIR setup (from 1000 to 1700~nm) only to the U-Ar and U-Ne lamps.\\

To characterize all lamps, we quantified the number of isolated metal and gas lines and analysed their spectral distribution. Another important characteristic of HCLs when used in astronomical spectrographs is that high intensity lines can produce saturation effects in the cgarge-coupled device (CCD). These lines are mostly emitted by the buffer gas of the HCLs. Therefore, the number of these high intensity lines and their spectral distribution should be identified.\\

\subsection{Analysis in the VIS}
\label{Ana_vis}

We systematically analysed the metal and gas lines of U-Ar, U-Ne, Th-Ar, and Th-Ne HCLs using the VIS setup. We averaged 200 scans per lamp taken under the same conditions (same imaging fore-optics and 12~mA operational current) at a resolution of 0.03~cm$^{-1}$. For our analysis, we identified the metal (U and Th) and buffer gas (Ne and Ar) lines emitted by all four HCLs as explained in Sect.~\ref{sec:lineide}. The number of confirmed metal and gas lines in all four lamps is summarized in Table~\ref{table_lineidentification_cathode_selection_notblended_vis}. In the last column of Table~\ref{table_lineidentification_cathode_selection_notblended_vis}, we show the number of "high intensity gas lines", which we define as the gas lines with intensities higher than the most intense metal line.\\
\begin{table}[h]
 \caption{Confirmed metal and gas lines using the VIS setup for the experiment \textit{cathode and buffer gas elements selection}. High intensity gas lines are defined as the gas lines with intensities higher than the most intense metal lines}% title of Table
\label{table_lineidentification_cathode_selection_notblended_vis}% is used to refer this table in the text
\centering  
\begin{tabular}{c c c c} 
\hline
\hline
\begin{tabular}{@{}c@{}}HCL \\ elements\end{tabular} & \begin{tabular}{@{}c@{}}Metal \\  lines\end{tabular}& \begin{tabular}{@{}c@{}}Gas\\ lines\end{tabular} & \begin{tabular}{@{}c@{}}High intensity \\ gas lines\end{tabular}\\
\hline           
U,Ne & 1899 & 235 & 43\\
U,Ar & 1998 & 195 & 26\\
Th,Ne & 938 & 231 & 66\\
Th,Ar & 1436 & 233 & 26\\
\hline
\end{tabular}
\end{table}

In the second column of Table~\ref{table_lineidentification_cathode_selection_notblended_vis}, we see a larger number of confirmed lines in the U HCLs than in the Th HCLs. The number of confirmed isolated U lines in the U-Ar spectrum (1998) is slightly larger than in the U-Ne spectrum (1899), but both numbers are still comparable. However, they are very different for the Th lamps: while we identify 1436 isolated Th lines in the Th-Ar lamp spectrum, we only find 938 isolated Th lines in the spectrum of the Th-Ne lamp. Comparing the Th-Ar and U-Ar lamps we observe the same number of confirmed high intensity Ar lines in both spectra. However, when we compare the number of confirmed high intensity Ne lines in the Th-Ne with the U-Ne lamps, we find significantly fewer strong Ne lines in the U-Ne lamp.\\

\begin{figure*}
 \centering
  \includegraphics[width=17cm]{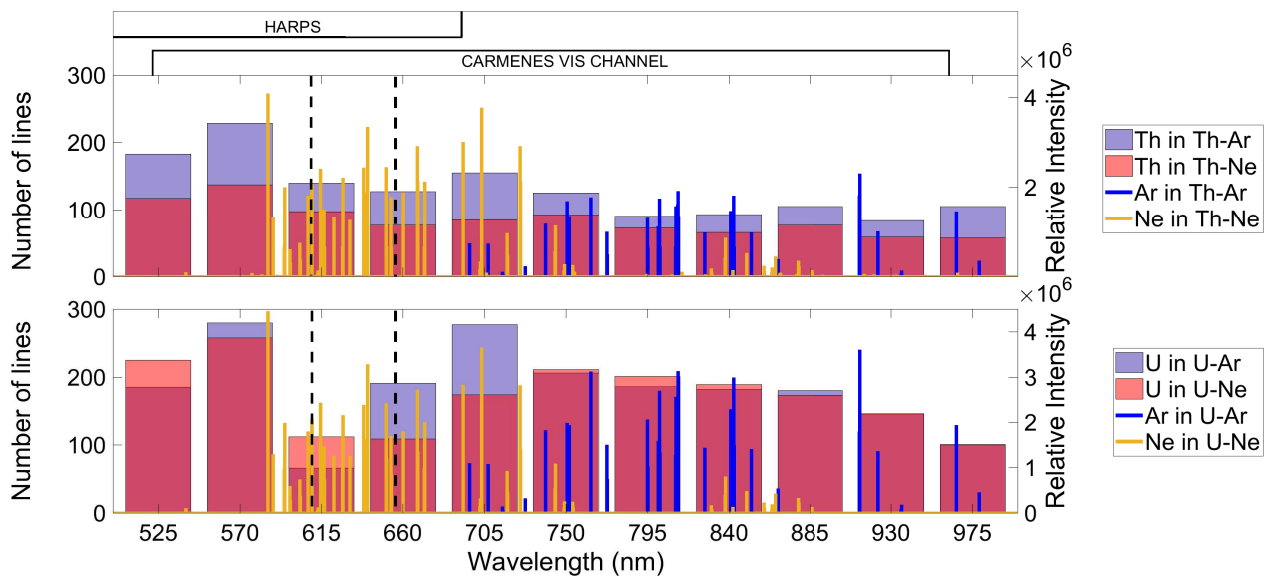}
   \caption{Histograms represent number of metal lines per 45~nm (light blue for Ar buffer gas HCLs and light red for Ne buffer gas lamps) using VIS setup. The area where both overlap is represented in magenta. The high intensity gas lines (i.e. gas lines with intensities higher than the most intense metal line) are overplotted. Argon lines are represented in blue and Ne lines in orange. In the top part, we indicate the spectral range covered by the CARMENES VIS channel and HARPS \citep{Mayor2003}. The spectral region between the dashed lines (from 610 to 655~nm) indicates where the internal laser of our FTS contaminates the recorded spectra. \textbf{Upper panel}: Th cathode HCLs (Th-Ne and Th-Ar). \textbf{Lower panel}: U cathode HCLs (U-Ne and U-Ar). Top part of the plot shows the spectral range coverage of high resolution astronomical 
spectrographs. The spectral region within the dashed lines indicates where the line identification is affected by the internal FTS laser (see Sect.~\ref{sec:Experimental setup and measurements}).}
      \label{test_thne_thar_VIS_uar_une_VIS_gas_above_maxU_NEW}
\end{figure*}

Besides the total number of metal lines, their spectral distribution is also important, as explained in Sect.~\ref{Int}. Therefore, we plot the histogram of the number of confirmed metal lines as a function of wavelength in Fig.~\ref{test_thne_thar_VIS_uar_une_VIS_gas_above_maxU_NEW} for all four lamps. We also indicate the confirmed high intensity gas lines as vertical lines.\\

We first discuss the upper panel of Fig.~\ref{test_thne_thar_VIS_uar_une_VIS_gas_above_maxU_NEW} by comparing the Th-Ar and Th-Ne lamps. We see that there are fewer confirmed Th lines in the Th-Ne lamp throughout the entire spectral range of the VIS setup. The difference in the number of confirmed Th lines between Th-Ar and Th-Ne is relatively homogeneously distributed. The number of confirmed gas lines, however, is comparable in both lamps ($233$ in Th-Ar and $231$ in Th-Ne) although the Ne lines show higher intensities than Ar lines. Twenty-six of the Ar lines in the Th-Ar lamp are high intensity gas lines. For the Th-Ne lamp we count $66$ high intensity gas lines, which is more than a factor of two higher than in the Th-Ar lamp. It is interesting to note that high intensity Ne lines are mainly emitted at wavelengths shorter than 750~nm, while the high intensity Ar lines are emitted at wavelengths longer than 695~nm. From the total number of Th lines and their spectral distribution, we conclude that Th-Ar 
is a better choice for wavelength calibration than Th-Ne for any spectrograph operating in the range from 500 to 1000~nm.\\

After analysing the Th cathode lamps, we take a close look at the lower panel of Fig.~\ref{test_thne_thar_VIS_uar_une_VIS_gas_above_maxU_NEW} and compare U-Ar with U-Ne. The number of confirmed Ne lines in the U-Ne spectrum (235) is slightly higher than the confirmed Ar lines (195) in the U-Ar lamp. Moreover, we find 43 high intensity Ne lines, which is a factor of 1.6 larger than the number of high intensity Ar lines (26). Most of the high intensity Ne lines are emitted at short wavelengths (shorter than 750~nm) while Ar lines are mostly emitted at wavelengths longer than 695~nm, as we previously saw for the Th cathode lamps. We identify a slightly larger number of confirmed metal lines and a lower number of high intensity lines and conclude that U-Ar performs better than U-Ne HCLs in the VIS band.\\

After comparing lamps with the same cathode material, our next goal is a comparison of different cathode materials, Th and U, regarding their usefulness for calibration purposes in the VIS spectral range (from 500 to 1000~nm). The position of brighter lines can be determined with higher accuracy, and a high number of calibration lines in the spectra is desirable for computing the wavelength solution of astronomical spectrographs. To account for both, we define the quality factor of a lamp as the sum of all line intensities in a given spectral range.\\

To compare lamps with this method, we normalized the spectra by the intensity of the gas lines because the intensity of these lines typically limits the exposure time of a detector. We calculate the normalizing factor of the Ar HCLs spectra using the 26 high intensity Ar lines identified in the U-Ar and Th-Ar spectra, and the normalizing factor of the Ne HCL spectra using the 43 high intensity Ne lines in the spectra of the U-Ne and Th-Ne HCLs. After normalizing the spectra, the total line intensity is computed over spectral ranges of 45 nm bins.\\

\begin{figure*}
 \centering
  \includegraphics[width=17cm]{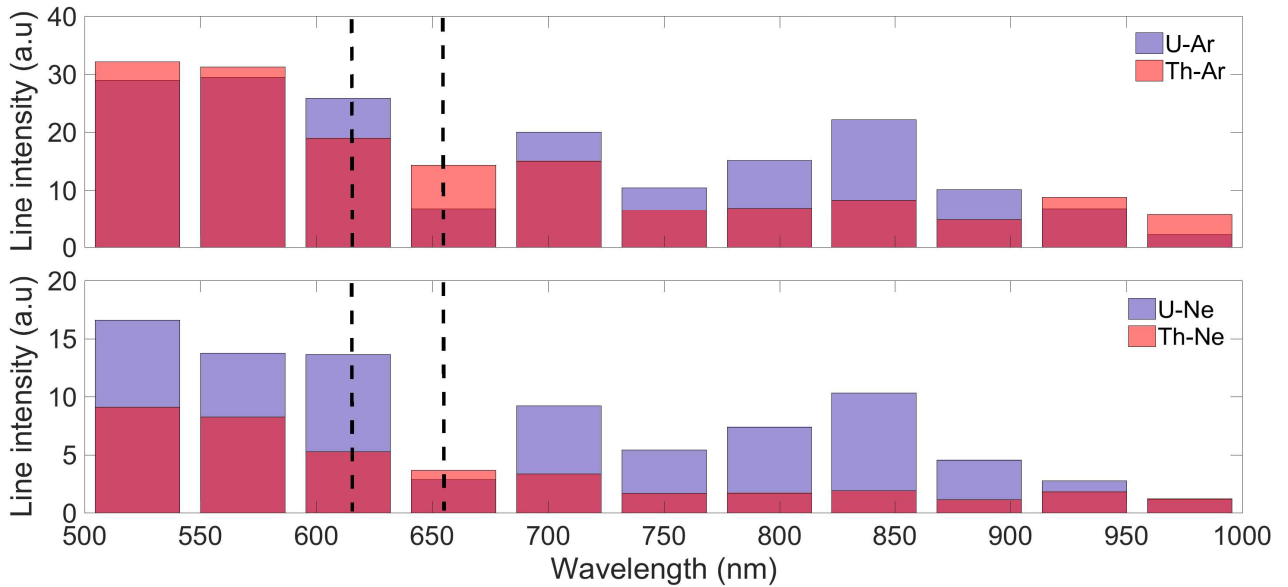}
   \caption{Total line intensity of metal lines per 45 nm in VIS setup. Blue bars represent the total line intensity per spectral range of U lines and red represents Th lines. Magenta indicates the overlap areas. The spectral region between the dashed lines (from 610 to 655~nm) indicates where the internal laser of our FTS contaminates the recorded spectra. \textbf{Upper panel:} HCLs with Ar buffer gas (U-Ar and Th-Ar). \textbf{Bottom panel:} HCLs with Ne buffer gas (U-Ne and Th-Ne).}
      \label{total_li_histo_TFC_DL}
\end{figure*}

In Fig.~\ref{total_li_histo_TFC_DL} the uppel panel depicts the results of the HCLs with Ar buffer gas and the bottom panel shows the results of the HCLs with Ne buffer gas. The results from U-Ar and Th-Ar are comparable in the first two wavelength bins from 500 to 591~nm and in the last two wavelength bins from 909 to 1000~nm. For the rest of the VIS spectra (with the exception of the central wavelength range from 636 to 681~nm), the U-Ar lamp clearly outperforms the Th-Ar lamp. This is also reflected in the total line intensity when summed over the entire VIS spectral range. Overall, the U lines in the U-Ar spectrum exhibit 16\% more calibration light than the Th lines in the Th-Ar spectrum. For the lamps with 
Ne buffer gas (bottom panel in Fig.~\ref{total_li_histo_TFC_DL}), we see that the total line intensity per spectral range is higher in almost the entire VIS spectral range for the U-Ne lamp. The only exception is the wavelength bin centred at 660~nm, where U-Ne and Th-Ne are comparable. The total amount of light available in the U lines of the U-Ne lamp is 222\% greater than the total amount of light available in the Th lines of the Th-Ne lamp.\\

From our results we conclude that, in the defined VIS spectral range (from 500 and 1000~nm), U HCLs are more suitable for wavelength calibration of high resolution spectrographs than Th HCLs. Among the U lamps, we identify a slightly higher number of confirmed isolated cathode lines in the U-Ar than in U-Ne spectrum. Although both lamps show a considerable number of filling gas lines, the high intensity gas lines are more numerous and also brighter in the U-Ne HCL. The high intensity lines can saturate the CCD of the spectrographs, which can contaminate extended regions of the spectrum and make them useless for wavelength calibration. Therefore, U-Ar lamps show better properties than U-Ne HCLs for wavelength calibration.\\

To improve the wavelength solution in the spectral regions of strong gas lines (Ar or Ne), one strategy is to use both the U-Ar and U-Ne lamp independently. Uranium lines blended with strong gas lines in one of the lamps will be uncontaminated in the other because Ne and Ar emit lines in different spectral regions. The U lines found in both spectra can then be combined and used for wavelength calibration. This strategy is currently used in the VIS channel of the CARMENES spectrograph.\\

\subsection{Analysis in the NIR}
\label{Ana_nir}

After analysing the VIS setup, we performed the same analysis for U-Ne and U-Ar HCLs in the NIR wavelength range (from 1000 to 1700~nm). We averaged 200 scans taken at a resolution of 0.02~cm$^{-1}$ while the lamps were operated at 12~mA. We started by identifying the U and buffer gas lines. The number of confirmed U and Ne lines, and the high intensity gas lines found in the two spectra, are summarized in Table~\ref{table_lineidentification_cathode_selection_notblended_nir}.\\
\begin{table}[h]
 \caption{As in Table~\ref{table_lineidentification_cathode_selection_notblended_vis}, but for U cathode lamps.}% title of Table
\label{table_lineidentification_cathode_selection_notblended_nir}      % is used to refer this table in the text
\centering  
\begin{tabular}{c c c c} 
\hline
\hline
\begin{tabular}{@{}c@{}}HCL \\ elements\end{tabular} & U& \begin{tabular}{@{}c@{}}Gas \\ lines\end{tabular} & \begin{tabular}{@{}c@{}}High intensity \\ gas lines\end{tabular}\\
\hline           
U,Ne & 1538 & 188 & 15\\
U,Ar & 1480 & 179 & 31\\
\hline
\end{tabular}
\end{table}

From Table~\ref{table_lineidentification_cathode_selection_notblended_nir}, we can see that the number of confirmed U lines in the U-Ne HCL (1538) is slightly higher than in the U-Ar lamp (1480). Although we identify a similar number of confirmed gas lines (188 Ne and 179 Ar lines), the number of high intensity Ar lines (31) is about a factor of two larger than the number of Ne lines (15).\\

\begin{figure*}
 \centering
  \includegraphics[width=17cm]{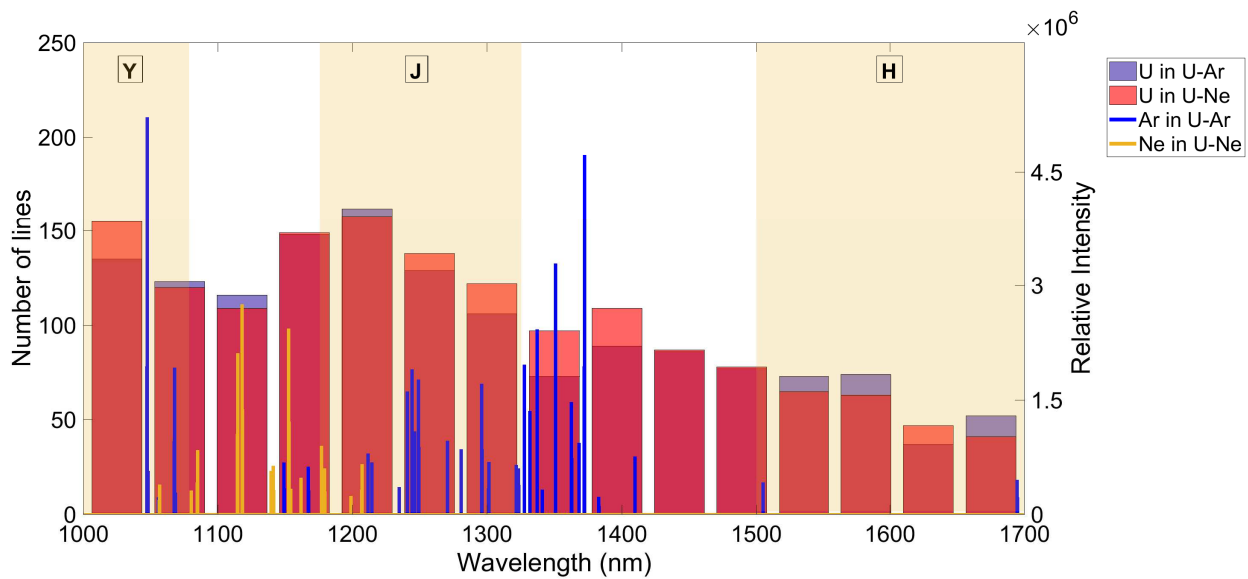}
   \caption{Same as in Fig.~\ref{test_thne_thar_VIS_uar_une_VIS_gas_above_maxU_NEW} but for NIR wavelengths where only U HCLs (U-Ne and U-Ar) have been studied. In light yellow we indicate the astronomical \textit{Y}, \textit{J} and \textit{H} bands.}
      \label{test_uar_une_NIR_gas_above_maxU_NEW}
\end{figure*}

We plot the histogram of the number of confirmed U lines as a function of wavelength in Fig.~\ref{test_uar_une_NIR_gas_above_maxU_NEW}. We overplot the high intensity lines emitted by the buffer gas of the lamps as vertical lines.\\ 

As seen in Fig.~\ref{test_uar_une_NIR_gas_above_maxU_NEW}, the metal line distributions of both lamps are fairly comparable. We can see a small dip of the line density around 1100 nm as a result of the lack of sensitivity in the instrument used to prepare the R2011 line-list that is used to identify U lines in this section (as explained in Sect.~\ref{Int}). It is interesting to note that the number of high intensity gas lines in the astronomical \textit{Y}, \textit{J,} and \textit{H} bands (indicated in light yellow in Fig.~\ref{test_uar_une_NIR_gas_above_maxU_NEW}) is higher in the U-Ar HCL than in the U-Ne.\\

Since the number of metal lines and their distribution are comparable, but the number of high intensity gas lines is lower in the U-Ne lamps, we conclude that for the spectral range between 1000 and 1700~nm, U-Ne HCLs are more suitable for wavelength calibration of high resolution astronomical spectrographs than U-Ar HCLs.\\
%______________________________________________________________ 
\section{Molecular contamination}
\label{sec:mol cont}

\begin{figure}[h]
 \centering
  \includegraphics[width=\hsize]{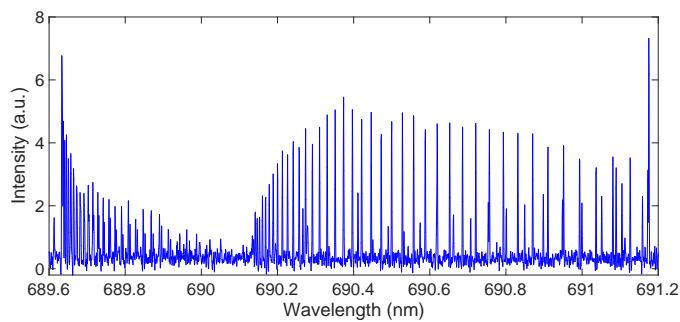}
   \caption{Example of molecular band observed in the Th-Ar HCL spectrum. See Table~\ref{table_molecular_bands} for a complete list.}
      \label{thar_mol_bands}
\end{figure}

In the analysis described in the Sect.~\ref{sec:cathode element selection} we explained that we did not detect significant molecular contamination from other species. Nevertheless, we did find seven molecular bands distributed from 612 to 947~nm of the Th-Ar lamp spectrum (see Table~\ref{table_molecular_bands}). These molecular bands are a source of noise for wavelength calibration. In the case of the Th-Ar HCL, they can contaminate up to 7.64~nm (e.g. molecular band identified from 897.42 to 905.06~nm) of the spectrum. In Fig.~\ref{thar_mol_bands}, the molecular band identified from 689.61 to 691.56~nm is shown as an example. We observed these features in the new Th-Ar HCLs; we did not find any similar features in our U lamps and Th-Ne HCL.\\
\begin{table}[h]
 \caption{Molecular bands emitted by the Th-Ar HCL identified in the spectrum recorded using the VIS setup. We define \textit{$\lambda_{s}$} and \textit{$\lambda_{e}$} as the start and end wavelength (measured in vacuum) of the molecular band.}             % title of Table
\label{table_molecular_bands}      % is used to refer this table in the text
\centering                          % used for centering table
\begin{tabular}{c c}        % centered columns (4 columns)
\hline\hline                 % inserts double horizontal lines
$\lambda_{s}$ (nm) & $\lambda_{e}$ (nm)\\% table heading 
\hline
612.31 & 612.48\\
689.60 & 691.56\\
693.05 & 696.53\\
727.64 & 728.71\\
787.03 & 791.64\\
897.42 & 905.06\\
942.31 & 946.52\\
\hline  
\end{tabular}
\end{table}

%______________________________________________________________ 
\section{Comparison of 12 U-Ne HCLs}
\label{sec:common and individual properties of the HCLs}

HCLs can show different properties depending on the manufacturer and the manufacturing procedure. The number of emission lines, line intensity ratio between metal and gas lines, and buffer gas contamination have an effect on the performance of HCLs as wavelength calibrators.  We analysed the consistency of these characteristics for 12 U-Ne HCLs with the NIR setup. All lamps are from the same batch and have consecutive serial numbers. For each lamp we co-added 120 individual scans at a resolution of 0.035~cm$^{-1}$. The lamps were operated at 6~mA resulting in lower SNR spectra than in the previous sections. We finely aligned each lamp before its measurement with a power-meter (see Sect.~\ref{sec:Experimental setup and measurements}).\\

First, we searched for U and Ne lines, and for gas contaminants in the lamps as explained in Subsection~\ref{sec:lineide}. We did not identify any significant gas contamination from other species in any of our spectra. Second, we selected the lines with intensities above a defined threshold (ten times the noise level). We summarize the results in Table~\ref{table_identified_lines_12_hcls}. We find that the number of U~{\sc i} lines differs considerably for different lamps (e.g. in the lamp 0127 we find a factor of 1.8 more lines than in the lamp 0120). The number of U~{\sc ii}, Ne~{\sc i,} and Ne~{\sc ii} lines is essentially constant.\\

\begin{table}[h]
 \caption{Number of confirmed isolated U and Ne lines with SNR higher than ten found in the 12 U-Ne spectra for NIR wavelengths. The last column contains the average of the U line intensity normalized to the average of its line intensity in the 12 spectra ($I^{U}$).}               % title of Table
\label{table_identified_lines_12_hcls}      % is used to refer this table in the text
\centering                          % used for centering table
\begin{tabular}{l c c c c c c}        % centered columns (4 columns)
\hline\hline                 % inserts double horizontal lines
\begin{tabular}{@{}c@{}} HCL \\ HKH\end{tabular} & U~{\sc i} & U~{\sc ii} & Ne~{\sc i} & Ne~{\sc ii} &  $I^{U}$\\% table heading 
\hline
0120 & 273 & 11 & 158 & 40 & 0.74\\
0121 & 377 & 12 & 151 & 39 & 1.08\\
0122 & 294 & 9 & 153 & 43 & 0.80\\
0123 & 334 & 12 & 158 & 44 & 0.95\\
0124 & 351 & 13 & 145 & 39 & 1.04\\
0125 & 307 & 9 & 162 & 40 & 0.82\\
0126 & 295 & 13 & 155 & 42 & 0.80\\
0127 & 503 & 11 & 153 & 40 & 1.39\\
0128 & 366 & 10 & 144 & 45 & 1.12\\
0129 & 398 & 12 & 154 & 39 & 1.09\\
0130 & 315 & 13 & 157 & 42 & 0.83\\
0131 & 478 & 11 & 143 & 41 & 1.35\\
\hline  
\end{tabular}
\end{table}

We looked for the confirmed lines that can be found in all of the 12 spectra to systematically analyse the difference between the 12 lamps. In total, we detected 222 U~{\sc i} and 104 Ne~{\sc i} lines in the spectra of all lamps. We normalized the 12 spectra to the the sum of the Ne lines intensities averaged in each spectrum. Then we divided the line intensity of each U line by the average line intensity of this line in the 12 spectra. Last, we calculated the average of the normalized line intensity of the 222 U~{\sc i} lines (named $I^{U}$ in Table~\ref{table_identified_lines_12_hcls}) for each lamp.\\
\begin{figure}[h]
 \centering
  \includegraphics[width=\hsize]{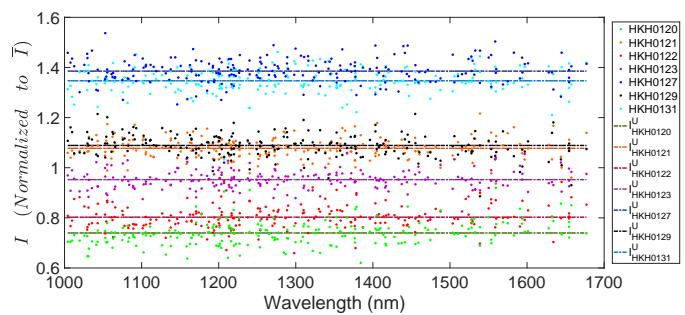}
   \caption{Coloured dots represent the normalized line intensity of 222 U~{\sc i} lines for each HCL in the spectral range from 1000 to 1700~nm. Dashed coloured lines indicate $I^{U}_{HKHi}$ of each lamp (i.e. average of the 222 U~{\sc i} normalized line intensity). For readability reasons we plot only seven of the 12 HCLs.}
      \label{normalize_SNR_to_average_SNR}
\end{figure}

In Fig.~\ref{normalize_SNR_to_average_SNR} we plot the normalized line intensity of the common U~{\sc i} lines in the NIR spectral range for seven HCLs (for readability reasons). The dashed lines represent $I^{U}$ for each HCL. From Fig.~\ref{normalize_SNR_to_average_SNR} and Table~\ref{table_identified_lines_12_hcls} it can clearly be seen that $I^{U}$ varies significantly between the lamps. Lamps with the lowest U line intensities are HKH0120 and HKH0122. Lamps with the highest U line intensities are HKH0127 and HKH131. The maximum U line intensity difference is 88\% (with respect to the minimum value), measured between lamps HKH0127 and HKH0120.\\

A possible explanation for the observed behaviour of the HCLs could be a different geometry of the lamps or in the light-fibre coupling efficiency, as seen in \citet{Huke}. This could affect the gas-metal intensity ratio. Another explanation could be that the metal line intensities are very sensitive to the internal pressure of the buffer gas \citep{Crosswhite55}. Therefore, even small changes in pressure from lamp to lamp would have an effect on the metal line intensity. We conclude that HCLs can show noticeable differences in number and intensity of metal lines even if they are manufactured in a single batch.\\

\section{U line-list}
\label{sec:U_line_list}
We concluded in Sect.~\ref{sec:cathode element selection} that U cathode lamps are more suitable for wavelength calibrators than Th lamps from 500 to 1700~nm. To use HCLs as wavelength calibrators for high precision RV spectrographs, the wavelength of the spectral lines must be known with high accuracy and precision. We present a new line-list for U (ranging from 500 to 1700~nm) that complements that of R2011.\\

\subsection{Line-list in the VIS}
\label{sec:linelist_vis}
The VIS setup is used to record spectra from 500 to 1000~nm of one U-Ne HCL operated at 6, 9, and 12~mA. The FTS is set to a resolution of 0.03~cm$^{-1}$ to record 1000 single scans that are co-added. We identified the confirmed isolated U and Ne lines as explained in Sect.~\ref{sec:Methods for calibration and analysis}. The detailed number of lines found in each step in the different spectra is presented in Table~\ref{table_lineidentification_linelist_vis} in Appendix A.\\ 

In Fig.~\ref{number_of_lines_vis}, we plot the number of confirmed U lines (i.e. U$_{I}$ and U$_{II}$), confirmed Ne lines (i.e. Ne$_{I}$ and Ne$_{II}$), and U line candidates, meaning that line candidates with FWHM comparable to the confirmed U lines but no corresponding line in the literature.\\ 

\begin{figure}[h]
 \centering
  \includegraphics[width=\hsize]{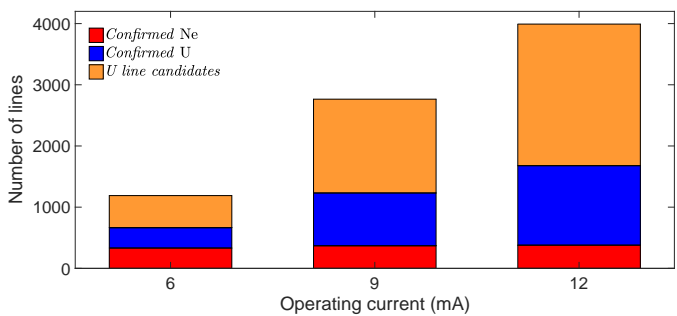}
   \caption{Number of lines in the spectra recorded when the U-Ne HCL was operated at 6, 9, and 12~mA using the VIS setup (from 500 to 1000~nm). The subset of confirmed U lines (with a corresponding line in the literature) is shown in blue. The subset of confirmed Ne lines is shown in red. We show the number of U line candidates in orange (line candidates with line width comparable to the FWHM of the confirmed U lines and with no match in the literature).}
      \label{number_of_lines_vis}
\end{figure}

In Fig.~\ref{number_of_lines_vis}, we see a large increase in the number of confirmed U lines for higher operational currents while the number of confirmed Ne lines remains almost constant. In order to identify a larger number of U lines, we looked at which of the detected lines have widths comparable to the width of those U lines that are already confirmed. We identify 2314, 1533, and 525 lines in the spectra recorded at 12, 9, and 6~mA respectively, and flag them as "U line candidates".\\

We carried out a second test to check whether our U line candidates are in fact emitted from the U cathode. If so, we expect the intensities of these lines to show a similar behaviour with increasing operational current as the confirmed U lines. We created four data subsets: the confirmed U lines with a line match in the 6, 9, and 12~mA spectra (258 lines); the confirmed Ne lines with a line match in the 6, 9, and 12~mA spectra (194 lines); the U line candidates found in the 6, 9, and 12~mA spectra (525 lines); and the U lines candidates found in 9 and 12~mA spectra only (1533 lines). Next, we normalized the line intensity of every line to its intensity at 9~mA. Lastly, we calculated the weighted average and the weighted standard deviation of the normalized line intensities for each data subset.\\

\begin{figure}[h]
 \centering
  \includegraphics[width=\hsize]{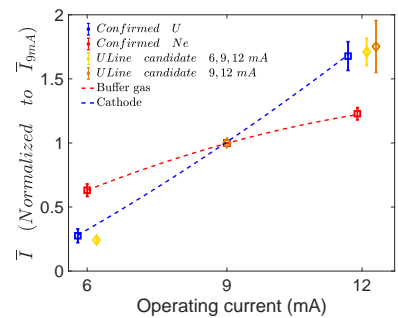}
   \caption{Weighted average of the line intensity normalized to the line intensity of the line in the 9~mA spectrum of the four data subsets: 258 confirmed U lines (blue squares) and 194 confirmed Ne lines (red squares), 525 U lines candidates with a match in the 6, 9, and 12~mA spectra (light brown diamonds), and 1533 U line candidates identified only in the 9 and 12~mA spectra (brown diamonds). The error bars indicate the weighted standard deviation. The red and blue dashed lines indicate the approximate trend that follows the normalized averaged intensity of the buffer gas and the cathode lines, respectively.}
      \label{visible_relativeI_opcurrent}
\end{figure}

In Fig.~\ref{visible_relativeI_opcurrent}, we plotted the weighted average and the weighted standard deviation of the four data subsets as a function of the operational current. First, we look at the line intensity of the confirmed U and Ne lines. We see that the line intensities of these subsets show a different growth rate for different operational currents (see Sect.~\ref{Int}). The line intensity of the U line candidates subsets is consistent with the behaviour of the confirmed U lines. Because both the FWHM of the U line candidates and their intensity-rise as a function of operational current are consistent with the confirmed U lines, we conclude that the emitting element of these 1533 lines is indeed U.\\

We provide an online line-list with the line position measured in vacuum, the line intensity, and the emitting element of all the confirmed and the U line candidates. Its format is explained in Table~\ref{line_list_vis}. In addition to the lines discussed above, we find $781$ lines that have a FWHM that is consistent with the FWHM of the confirmed U lines. We included these $781$ lines in the line-list with the flag '\textit{12}'. We also find six lines with FWHM consistent with the width of the confirmed U lines without a corresponding Ritz wavelength. The R2011 line-list also contains these lines (where they are flagged as \textit{'?'}). We checked their line intensity-rise with different operational currents. They show a consistent behaviour with the confirmed U lines. We included these six lines in our line-list and flagged them as '\textit{R2011?}'.\\

\begin{table}[h]
 \caption{Description of the line-list.} % title of Table
\label{line_list_vis}      % is used to refer this table in the text
\centering                          % used for centering table
\begin{tabular}{c c c l}        % centered columns (4 columns)
\hline\hline                 % inserts double horizontal lines
Column &  Symbol & Unit & Explanation\\% table heading 
\hline
1 & $\tilde{\nu}_{c}$ & cm$^{-1}$ & Line centre in wavenumber\\
2 & $\sigma_{\tilde{\nu}_{c}}$ & cm$^{-1}$ & Uncertainty of the line centre\\
3 & $\lambda$ & nm & Line centre in wavelength\\
4 & $\sigma_{\lambda}$ & nm & Uncertainty of the line centre\\
5 & I & au & Relative line intensity\\
6 & Species &  & Emitting element\\
7 & Notes & & 'R2011': contained in R2011\\
 & & & line-list.\\
 & & & 'ULC': U line candidate.\\
 & & & '$12$': U line candidate only\\
 & & & at 12~mA spectrum.\\
 & & & 'R2011?': U line candidate\\
 & & & (\textit{'?'} in R2011 line-list).\\
\hline  
\end{tabular}
\end{table}

\begin{figure}[h]
 \centering
  \includegraphics[width=\hsize]{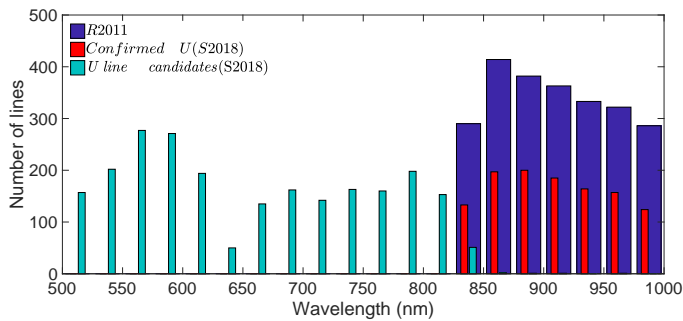}
   \caption{Uranium line density in the spectral range from 500 to 1000 nm. The blue portion denotes the line-list from the literature (R2011). We overplot the confirmed U lines of our work (S2018) in light red. The U line candidates are denoted in green. The bin size in this plot is constant at 25 nm.}
      \label{hisvis_redman_sarmi}
\end{figure}

We analysed the spectral distribution of the U lines. A uniform distribution of lines is desired when choosing a calibrator lamp for astronomical spectrographs. We plot in Fig.~\ref{hisvis_redman_sarmi} the histogram for the number of lines as a function of wavelength of the R2011 line-list (blue) and overplot the confirmed U lines in our work (Sarmiento et al. 2018, from now on S2018) contained in R2011 (red) and the U line candidates (green). The shortest wavelength in the R2011 line-list is 833~nm, which is the reason why we identify U line candidates bluewards of this wavelength.\\

When we compare the distribution of lines in the spectral region where our FTS setup overlaps with that of R2011 (blue and red histograms in Fig.~\ref{hisvis_redman_sarmi}), we observe that about $50$~\% of the lines listed in R2011 are identified in our data. We attribute this difference in the number of lines to the different operational current used in this work and R2011. While in R2011 the spectra were taken with operational currents between $26$ and $300$~mA, we used spectra taken at $6$, $9,$ and $12$~mA and therefore our line-list is limited to the brighter spectral features.\\

The number of identified lines in our spectra (confirmed $+$ U line candidates) decreases towards the blue and red end of the VIS spectral range. We attribute this decrease in number of lines to the spectral response sensitivity of our setup and instrument. In addition, we also observe a low number of lines in the region from 600 to 700 nm. This spectral band is strongly contaminated by the internal FTS He-Ne laser (see Sect.~\ref{sec:Experimental setup and measurements}), which is why the weaker spectral lines are likely missed.\\

The spectrum of the U lamp does not show any extended spectral region that is completely devoid of U lines. Overall, the number of spectral lines per $25$~nm bin is relatively constant and varies only up to a factor of two in our line-list (except for the bin most contaminated by the He-Ne laser centred on $637.5$~nm). The overall number of calibration lines available and their relatively uniform spectral distribution make U lamps good candidates to calibrate astronomical spectrographs operating in optical wavelength regimes.\\

\begin{figure}[h]
 \centering
  \includegraphics[width=\hsize]{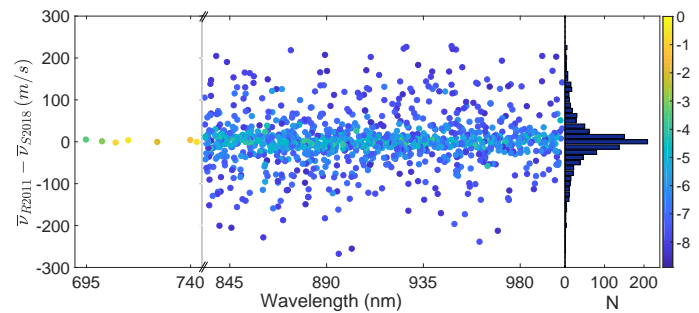}
   \caption{\textbf{Left panel:} Line position difference in velocity units of the isolated confirmed U lines (i.e. with a corresponding match in the literature). The colour scale located at the right edge of the plot denotes the normalized line intensity in a logarithmic scale. From 690 to 744~nm, we can see the residuals of the eight standard lines used to calculate the $\overline{k}_{VIS}$. The x-axis is shortened for readability reasons at 744~nm. \textbf{Right panel:} Distribution of the residuals. The distribution is centred on $-0.12 \pm 8.19${m/s}$^{-1}$, where the uncertainty is calculated as the weighted standard deviation plus the statistical error introduced by the standard lines used for calibration.}
      \label{Red_IAG_overlaped_vis}
\end{figure}

We investigated possible systematics in the wavelength positions of our line-list using the R2011 line-list in the overlapping spectral region from 833 to 1000 nm. The left panel of Fig.~\ref{Red_IAG_overlaped_vis} shows the difference in line position between R2011 and our work; the histogram in the right panel shows the distribution of the residuals as a function of velocity units.\\ 

In Fig.~\ref{Red_IAG_overlaped_vis} we do not see systematics in our work when comparing with the R2011 line-list. The mean difference between the two line-lists is $-0.12 \pm 8.19{m/s}$ (weighted mean and standard deviation). We conclude that the line positions in our line-list are consistent with those in R2011 within the error bars.\\

Our VIS line-list provides a total of $3480$ U lines, of which $2320$ lines are U line candidates. There are $2314$ lines that are not contained in R2011 because the line-list does not show information in wavelengths bluewards of $833$~nm (plus six U line candidates that are flagged as \textit{'R2011?'}). We conclude that the even distribution and high abundance of lines render U lamps excellent calibrators for both NIR and VIS range spectrographs.\\

\subsection{Line-list in the NIR}
\label{sec:linelist_nir}

\begin{figure}[h]
 \centering
  \includegraphics[width=\hsize]{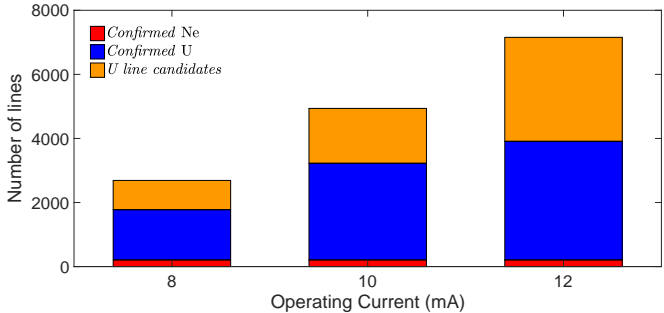}
   \caption{Same as in Fig.~\ref{number_of_lines_vis}, but for NIR wavelengths (from 1000 to 1700 nm), when the U-Ne HCL was operated at 8, 10, and 12~mA.}
      \label{number_of_lines_nir}
\end{figure}

We recorded and co-added 450 single scans of the U-Ne lamp at a resolution of 0.01~cm$^{-1}$, using the NIR setup from 1000~nm to 1700~nm (10000~cm$^{-1}$ to 5882.4~cm$^{-1}$). The U-Ne HCL was operated at three different operational currents 8, 10, and 12~mA (see Table~\ref{table_experiments} for further details). We detected and identified the lines as explained in Sect.~\ref{sec:Methods for calibration and analysis}. In Fig.~\ref{number_of_lines_nir}, we plot the number of confirmed lines and U line candidates in the NIR spectra.\\  

\begin{figure}[h]
 \centering
  \includegraphics[width=\hsize]{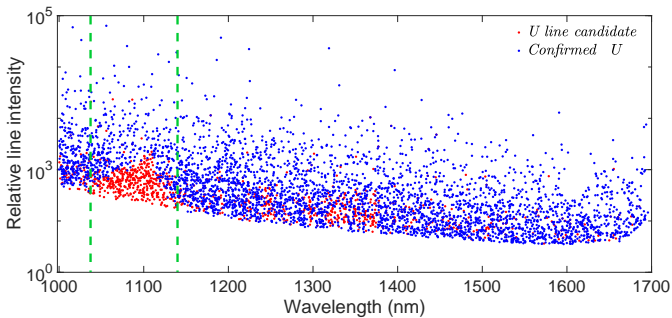}
   \caption{Relative line intensity of the U line candidates (red dots) and the confirmed U lines (blue dots) as a function of wavelength. The green dashed lines indicate the spectral band from 1031 to 1139~nm, where the R2011 line-list shows a lack of emision lines caused by the spectral response of its spectrograph.}
      \label{lineidendetect}
\end{figure}

As in the VIS setup, we can see in Fig.~\ref{number_of_lines_nir} that the number of confirmed U lines increases substantially at higher operational currents, while the number of confirmed Ne lines remains constant (see Sect.~\ref{Int}). We see a large number of U line candidates in the three spectra. Since our NIR setup overlaps with the spectral range of R2011, we compared the line position and intensity of our U line candidates and the confirmed U lines (i.e. with a match in R2011 line-list). In Fig.~\ref{lineidendetect} we plot the relative intensity of the U line candidates (red dots) and overplot the confirmed U lines (blue dots) found in both the 10~mA and the 12~mA spectra.\\

Two spectral regions in Fig.~\ref{lineidendetect} are of special interest: (i) in the range from 1031 to 1139~nm (indicated by the
green dashed lines), the R2011 line-list is lacking a number of lines that we found in our spectra, and (ii) there are several narrow
bands between 1200 and 1400\,nm where our spectra show U line candidates that are close to strong Ar lines (see Fig.~\ref{test_uar_une_NIR_gas_above_maxU_NEW}).\\

\begin{figure}[h]
 \centering
  \includegraphics[width=\hsize]{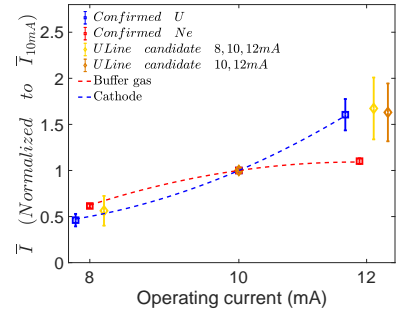}
     \caption{Same as in Fig.~\ref{visible_relativeI_opcurrent}, but when the U-Ne lamp is operated at 8, 10, and 12~mA and the line intensity is normalized to its intensity at 10~mA.}
      \label{lineintensity_diffopcurrent}
\end{figure}

We carried out a second test to determine the emitting element of our U line candidates following the methodology indicated in Sect.~\ref{sec:linelist_vis}. We first created four data subsets; 1418 confirmed U lines found in all spectra (blue squares); 132 confirmed Ne lines contained in all spectra (red squares); 137 U line candidates found in all spectra (light brown diamonds); and 908 U line candidates identified in the spectra of 10 and 12~mA (brown diamonds). We normalized the intensity of every line to its intensity at 10\,mA, and we calculated the weighted average of the normalized line intensity of each data subset. We plot the results in Fig.~\ref{lineintensity_diffopcurrent}.\\

Figure~\ref{lineintensity_diffopcurrent} shows that the line intensities of both the confirmed U and Ne lines (blue and red squares, respectively) show different behaviour for different operational currents. We find that the growth in line intensity of both sets of U line candidates is consistent with the behaviour of the confirmed U lines. Because the FWHM of the U line candidates is also consistent with the width of the confirmed U lines, we conclude that these 908 lines are U lines. We also find 151 lines with FWHM consistent with the width of the confirmed U lines (flagged as \textit{'?'} in R2011). For these lines, no corresponding Ritz wavelength is available in R2011. We confirm that these are U lines because they show the same behaviour with operational currents as the confirmed U lines with Ritz wavelengths. We included these 151 lines in our line-list with the flag '\textit{R2011?}'.\\

\begin{figure}[h]
 \centering
  \includegraphics[width=\hsize]{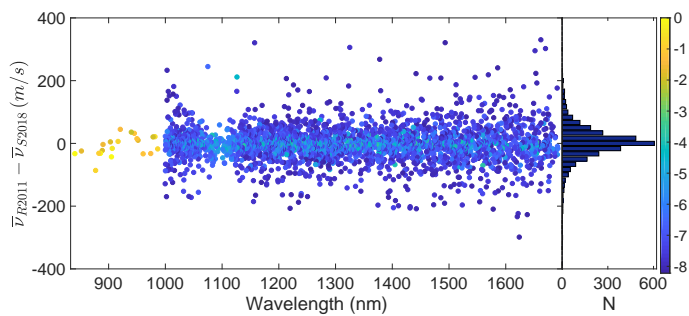}
   \caption{Same as Fig.~\ref{Red_IAG_overlaped_vis} but from 833 to 1700 nm. \textbf{Left panel:} The wavelength of the 21 lines used to calculate $\overline{k}_{NIR}$ factor are from 833 to 1000~nm. \textbf{Right panel:} The distribution is centred on $-3.5 \pm 11.0${m/s}$^{-1}$}
      \label{scat_histo_nir}
\end{figure}

We investigated possible systematics introduced by our NIR setup using the R2011 line-list. The left panel of Fig.~\ref{scat_histo_nir}
shows the difference in line position between our S2018 and the R2011 line-lists. There are no obvious systematic differences between the
two lists; higher intensity lines show less scatter than the lower intensity lines. In the right panel of Fig.~\ref{scat_histo_nir}, the
histogram shows the distribution of the differences. The mean difference between the two line-lists is $-3.5 \pm 11.0$\,m\,s$^{-1}$
(weighted mean and standard deviation), that is, the two distributions are consistent within the uncertainties.\\

\begin{figure}[h]
 \centering
  \includegraphics[width=\hsize]{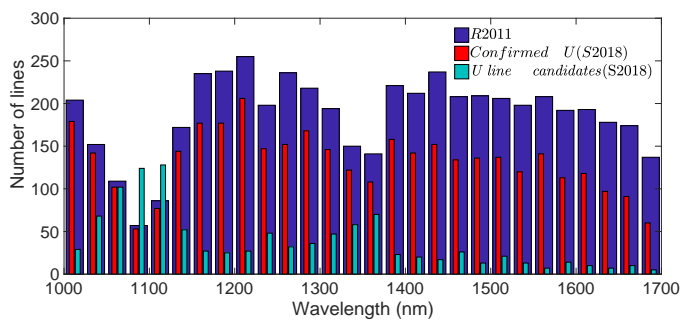}
   \caption{Same as Fig.~\ref{hisvis_redman_sarmi}, but for NIR wavelengths from 1000 to 1700~nm.}
      \label{histogram_paper_ulinesiagvsredman}
\end{figure}

The spectral distribution of U lines in our line-list obtained with the NIR setup is shown in Fig.~\ref{histogram_paper_ulinesiagvsredman}
with U lines of the R2011 line-list (blue), of our confirmed U lines (red), and of our U line candidates (green). The U lines are distributed across the entire NIR wavelength range rather homogeneously. Two spectral regions stand out in Fig.~\ref{histogram_paper_ulinesiagvsredman} (see also Fig.~\ref{lineidendetect}): (i) from 1031 to 1139~nm, the R2011 line list contains relatively few lines; and (ii) the spectrum is contaminated by high intensity Ar lines (see Fig.~\ref{test_uar_une_NIR_gas_above_maxU_NEW}). In the spectral region from 1031 to 1139~nm, our line-list almost doubles the number of known U lines: in R2011, 424 U lines are listed between 1031 to 1139~nm. While we identify 847 lines in this band, 444 of them U
line candidates. We see a decrease in the number of lines in the spectral range around 1230~nm and around 1340~nm in the R2011
line-list, where we identify 86 U and 218 U line candidates, respectively.

\section{Summary}

This work presents the analysis of four HCLs with different
combinations of cathode material and buffer gas. We recorded spectra
of U-Ne, U-Ar, Th-Ne, and Th-Ar lamps, compared their performance, and
created a new U line-list for the calibration of high resolution VIS
and NIR spectrographs.

We found that in the VIS, the U lamps exhibited a significantly larger
number of spectral lines than the Th lamps. For the lamps filled with
Ar, the difference was about 30\,\%. The difference was more severe in
lamps filled with Ne. Here, the U cathode provided about a factor of
two more lines than the Th cathode. The total number of lines was
comparable for the two U cathode lamps; the one filled with Ar showed
about 5\,\% more lines in the VIS wavelength range. This difference
was very large for the Th lamps where the lamp filled with Ar had
about 50\,\% more lines than the one filled with Ne.

Our analysis of the spectral line distribution demonstrated that the
combination of a U-Ne lamp and a U-Ar lamp can map out the wavelength
range of VIS spectrographs very efficiently. This strategy can provide
a high number of well defined cathode lines and mitigate the effect
of the pixel saturation because the strong gas lines from Ar and Ne
contaminate different wavelength ranges.

At NIR wavelengths, we only investigated U cathode lamps because Th
lamps are known to show fewer lines here. The U-Ne and U-Ar lamps
showed comparable performances that are very useful for wavelength
calibration. The Ne lamp spectrum showed fewer strong lines and might
therefore be preferred over the lamps filled with Ar.

As a product of our analysis, we presented a line-list for U in the
wavelength range 500--1700~nm. The catalogue contains 8239 unblended
lines that we attributed to U. The line-list allows the calibration of
high resolution spectrographs with U cathode lamps. It can be expected
to improve the wavelength calibration accuracy because of the larger
number of lines, and it provides an alternative to Th lamps also for
visual light spectrographs.

\begin{acknowledgements}
  LFS acknowledges support from the European Research Council under 
  FP7 GA 279347 and AY2011-30147-C03-03.\\
\end{acknowledgements}

%-------------------------------------------------------------------

\bibliographystyle{bibtex/aa} % style aa.bst
\bibliography{bibtex/reference} % your references Yourfile.bib
%\input{HCL_U_Th_Ull_S2018.bbl}

%\newpage
\begin{appendix} %First online appendix
\label{appendix_a}
\section{Summary of the different measurements and complementary figures.}
\begin{table*}[h]
 \caption{Table of the different measurements.}          % title of Table
\label{table_experiments}      % is used to refer this table in the text
\centering                          % used for centering table
\begin{tabular}{l c c c c c c c}        % centered columns (4 columns)
\hline\hline                 % inserts double horizontal lines
Experiment & \begin{tabular}{@{}c@{}}HCL \\ elements\end{tabular} & \begin{tabular}{@{}c@{}}Operational \\ current (mA)\end{tabular} & \begin{tabular}{@{}c@{}}Resolution  \\ (cm$^{-1}$)\end{tabular} &  \begin{tabular}{@{}c@{}}Scanning \\ Time (h)\end{tabular} & \# Scans & \begin{tabular}{@{}c@{}}Spectral \\ range\end{tabular}\\% table heading 
\hline
Properties of the U HCLs & U,Ne & 6 & 0.035 & 1.9 & 120 & NIR\\
\hline
Cathode selection & U,Ne & 12 & 0.03 & 2.7 & 200 & Vis\\
         & U,Ar & 12 & 0.03 & 2.7 & 200 & Vis\\
         & Th,Ne & 12 & 0.03 & 2.7 & 200 & Vis\\
         & Th,Ar & 12 & 0.03 & 2.7 & 200 & Vis\\
         & U,Ne & 12 & 0.02 & 3.3 & 200 & NIR\\
         & U,Ar & 12 & 0.02 & 3.3 & 200 & NIR\\
\hline  
Line-list & U,Ne & 6 & 0.03 & 13.7 & 1000 & Vis\\
         & U,Ne & 9 & 0.03 & 13.7 & 1000 & Vis\\
         & U,Ne & 12 & 0.03 & 13.7 & 1000 & Vis\\
         & U,Ne & 8 & 0.01 & 22.6 & 450 & NIR\\
         & U,Ne & 10 & 0.01 & 22.6 & 450 & NIR\\
         & U,Ne & 12 & 0.01 & 22.6 & 450 & NIR\\
\hline  
\end{tabular}
\end{table*}
\begin{table*}[h]
 \caption{Summary of the line identification for the experiment: cathode selection.}              % title of Table
\label{table_lineidentification_cathode_selection}      % is used to refer this table in the text
\centering  
\begin{tabular}{l c c c c c c c } 
\hline
\hline
Experiment & \begin{tabular}{@{}c@{}}HCL \\ elements\end{tabular} & \begin{tabular}{@{}c@{}}\# \textit{Detected} \\ \textit{lines}\end{tabular} & \begin{tabular}{@{}c@{}}Metal \\ lines\end{tabular} &  \begin{tabular}{@{}c@{}}Not blended \\ metal lines\end{tabular}& \begin{tabular}{@{}c@{}}Gas \\ lines\end{tabular}& \begin{tabular}{@{}c@{}}Not blended \\ gas lines\end{tabular}  & \begin{tabular}{@{}c@{}}Spectral \\ range\end{tabular}\\
\hline           
Cathode and & U,Ne & 8690 & 2256 & 1899 & 274 & 235 & Vis\\
buffer gas  & U,Ar & 9227 & 2374 & 1998 & 240 & 195 & Vis\\
selection   & Th,Ne & 6001 & 1044 & 938 & 270 & 231 & Vis\\
            & Th,Ar & 8323 & 1641 & 1436 & 275 & 233 & Vis\\
            & U,Ne & 4901 & 2080 & 1538 & 351 & 188 & NIR\\
            & U,Ar & 7431 & 1943 & 1480 & 315 & 179 & NIR\\
\hline
\end{tabular}
\end{table*}

\begin{table*}[h]
\caption{Line identification for the experiment: line-list in the VIS spectral range.}          % title of Table
\label{table_lineidentification_linelist_vis}      % is used to refer this table in the text 
\centering
\begin{tabular}{c c c c c c c c c c c}
\hline
\hline
\begin{tabular}{@{}c@{}}Operational \\ current (mA)\end{tabular} & \begin{tabular}{@{}c@{}}\# \textit{Detected} \\ \textit{lines}\end{tabular} & U$_{I}$ & U$_{I}^{Iso}$ & U$_{II}$ & U$_{II}^{Iso}$ & Ne$_{I}$ & Ne$_{I}^{Iso}$ & Ne$_{II}$ & Ne$_{II}^{Iso}$ & \textit{U$_{Candidates}$}\\
\hline
12 & 12378 & 1268 & 1132 & 31 & 28 & 248 & 229 & 131 & 124 & 2314\\
9 & 11602 & 845 & 749 & 18 & 16 & 244 & 225 & 125 & 110 & 1533\\
6 & 8969 & 329 & 274 & 3 & 2 & 232 & 212 & 100 & 75 & 525\\
\hline
\end{tabular}
\end{table*}

\begin{table*}[!htbp]
\caption{Line identification for the experiment: line-list in the NIR spectral range.}          % title of Table
\label{table_lineidentification_linelist_nir}      % is used to refer this table in the text 
\centering
\begin{tabular}{c c c c c c c c c c c}
\hline
\hline
\begin{tabular}{@{}c@{}}Operational \\ current (mA)\end{tabular} & \begin{tabular}{@{}c@{}}\# \textit{Detected} \\ \textit{lines}\end{tabular} & U$_{I}$ & U$_{I}^{Iso}$ & U$_{II}$ & U$_{II}^{Iso}$ & Ne$_{I}$ & Ne$_{I}^{Iso}$ & Ne$_{II}$ & Ne$_{II}^{Iso}$ & \textit{U$_{Candidates}$}\\
\hline
12 & 8723 & 3982 & 3550 & 169 & 150 & 192 & 160 & 62 & 49 & 3246\\
10 & 6211 & 3269 & 2942 & 81 & 74 & 191 & 160 & 62 & 49 & 1713\\
8 & 3306 & 1731 & 1549 & 19 & 17 & 189 & 157 & 52 & 39 & 916\\
\hline
\end{tabular}
\end{table*}

\begin{figure}[h]
 \centering
  \includegraphics[width=\hsize]{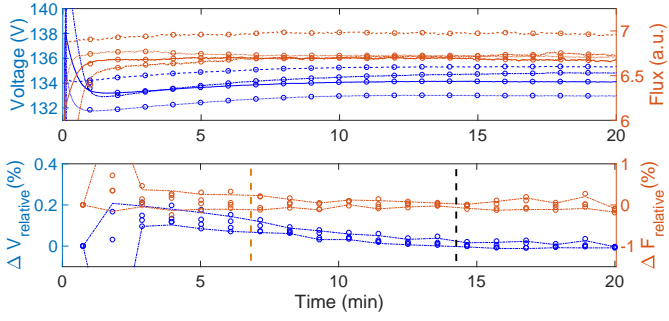}
   \caption{Voltage and flux measurement during warm-up time of the four different HCLs.\
   \textbf{Upper panel}: The blue lines indicate voltage measurements (y-axis in the left) and the brown lines represent the flux (y-axis in the right). Blue and brown circles indicate the one minute average values.\
   \textbf{Bottom panel}: The left axis indicates the relative voltage variation, and the right axis the relative flux variation. Blue and brown circles represent voltage and flux respectively. Blue and brown lines indicate $\pm \sigma$ of the calculated relative variation for voltage and flux. The brown dashed vertical line (at about seven minutes) indicates when the flux reaches a stability of 0.5\% for all the lamps. The black vertical dashed line indicates when the voltage is stable with relative changes lower than 0.05\%. We consider that the lamps offer their most stable output after 14-15 minutes.}
      \label{voltage_apendix}
\end{figure}
\begin{figure}[h]
 \centering
  \includegraphics[width=\hsize]{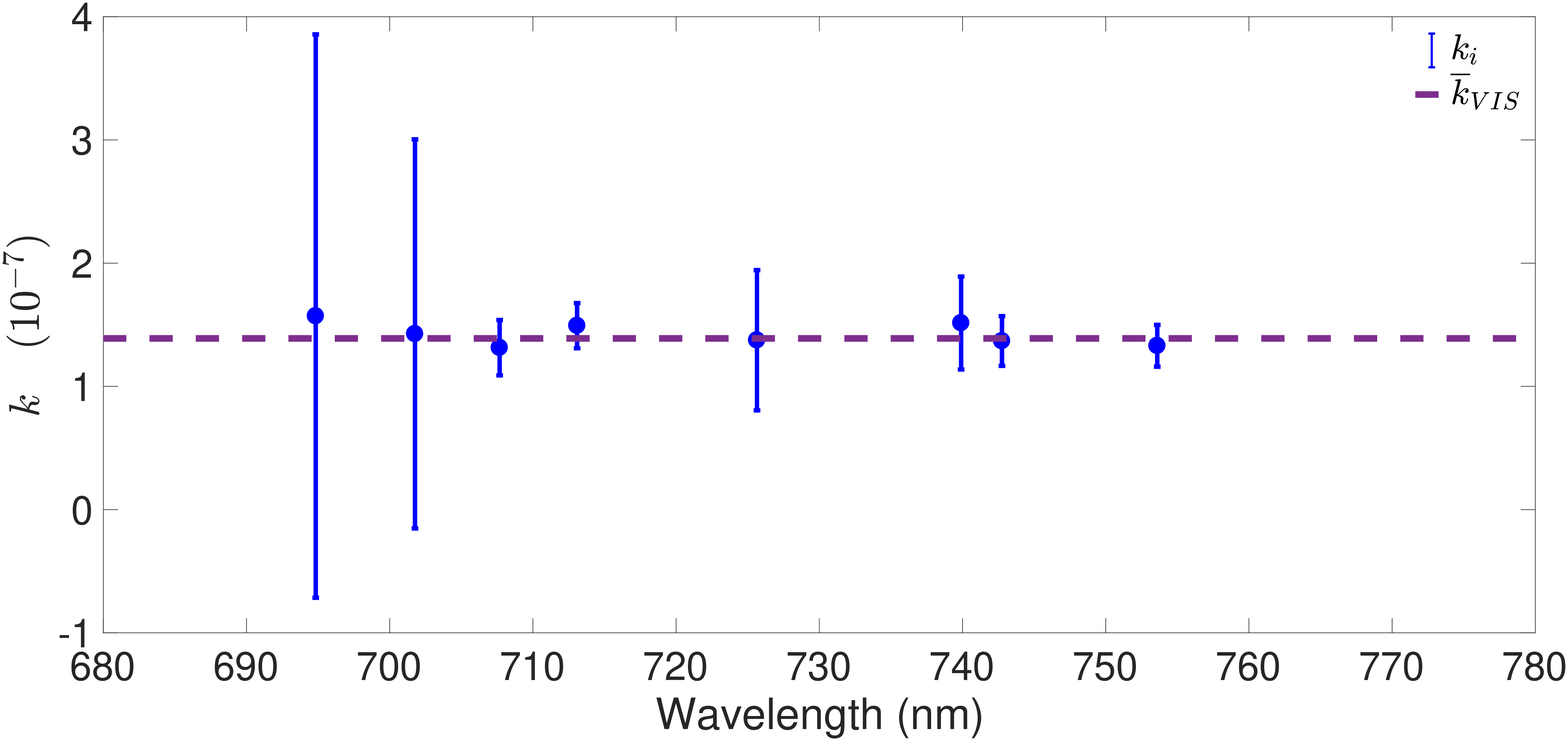}
   \caption{FTS \textit{k} factor used to obtain an absolute wavelength calibration of the spectra recorded from 500 to 1000 nm. We use eight U standard lines from \citet{DeGraffenreid2002} that have a match in our spectrum. The blue dots indicate the $k_{i}$ of each line. The error bars indicate the uncertainty calculated by error propagation of the line position measurement of our lines and the standard lines. The dashed purple line indicates the weighted average of the 8 $k_{i}$.}
      \label{k_vis}
\end{figure}

\begin{figure}[h]
 \centering
  \includegraphics[width=\hsize]{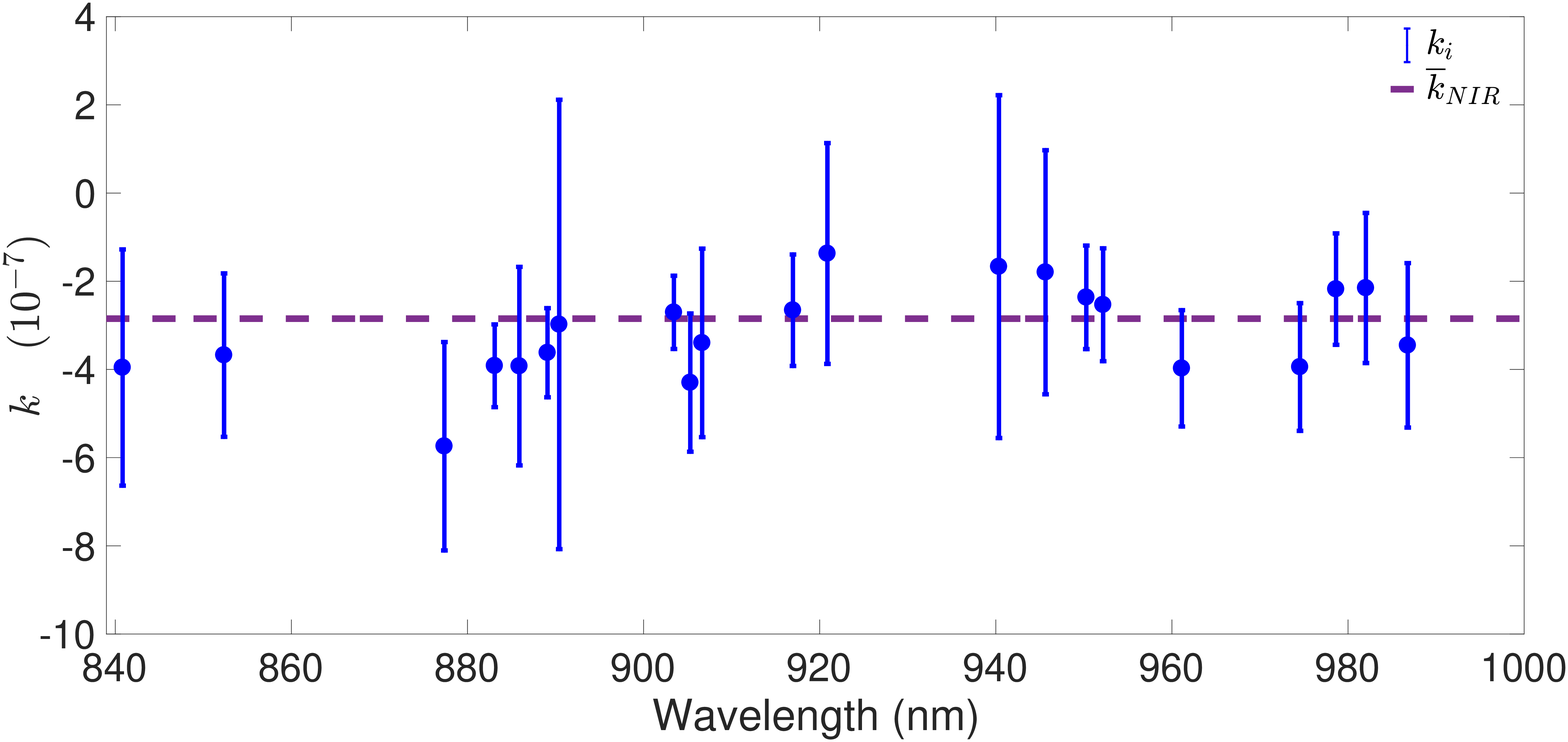}
   \caption{As in Fig.~\ref{k_vis}, but for the NIR spectrum. We use 21 high intensity isolated lines to calculate each $k_{i}$ confirmed in the VIS and in the NIR spectrum.}
      \label{k_nir}
\end{figure}

\end{appendix}
\end{document}